\documentclass[12pt]{article}
\usepackage{epsfig}
\newcommand{\betavar}{{\beta}_{\varphi}}
\newcommand{\betaphi}{{\beta}_{\phi}}
\newcommand{\betachi}{{\beta}_{\chi}}
\newcommand{\tbetavar}{{\tilde\beta}_{\varphi}}
\newcommand{\tbetaphi}{{\tilde\beta}_{\phi}}
\newcommand{\tbetachi}{{\tilde\beta}_{\chi}}
\newcommand{\Gvar}{G_{\varphi}}
\newcommand{\Gphi}{G_{\phi}}
\newcommand{\Gchi}{G_{\chi}}
\newcommand{\tGvar}{{\tilde G}_{\varphi}}
\newcommand{\tGphi}{{\tilde G}_{\phi}}
\newcommand{\tGchi}{{\tilde G}_{\chi}}
\newcommand{\gammam}{{\gamma}_{m}}
\newcommand{\gammaL}{{\gamma}_{\Lambda}}
\newcommand{\gammapsi}{{\gamma}_{\psi}}
\newcommand{\tgammam}{{\tilde\gamma}_{m}}
\newcommand{\tgammaL}{{\tilde\gamma}_{\Lambda}}
\newcommand{\tgammapsi}{{\tilde\gamma}_{\psi}}
\newcommand{\tpsi}{{\tilde\psi}}
\newcommand{\tL}{{\tilde\Lambda }}
\newcommand{\tm}{{\tilde m }}
\newcommand{\tM}{{\tilde M }}
\begin{document}
%
%  Title Page
%
\begin{titlepage}
\title{
  RENORMALIZATION GROUP AND RELATIONS
  BETWEEN SCATTERING AMPLITUDES IN A THEORY WITH
  DIFFERENT MASS SCALES}

\author{
A.V.Gulov
\thanks{e-mail address: gulov@ff.dsu.dp.ua}
~and V.V.Skalozub
\thanks{e-mail address: skalozub@ff.dsu.dp.ua}\\
\\
{\em Dniepropetrovsk State University} \\
{\em Dniepropetrovsk, 320625 Ukraine }
}
\date{November 2, 1998}
\maketitle
\thispagestyle{empty}
%     Abstract
%
\begin{center}
\begin{abstract}%=======================================================
In the Yukawa model with two different mass scales the
renormalization group equation is used to obtain relations between
scattering amplitudes at low energies. Considering fermion-fermion
scattering as an example, a basic one-loop renormalization group
relation is derived which gives possibility to reduce the problem
to the scattering of light particles on the "external field"
substituting a heavy virtual state. Applications of the results to
problems of searching new physics beyond the Standard Model are
discussed.
\end{abstract}%=========================================================
\end{center}
%
%  End of title page
%
\end{titlepage}

\section{Introduction}

An important problem of nowadays high energy physics is searching
for deviation from the Standard Model (SM) of elementary particles
which may appear due to heavy virtual states entering the extended
models and having the masses much greater than the W-boson mass
$m_W$ \cite{1}. One of approaches for the description of such
phenomena is the construction of the effective Lagrangians (EL)
appearing owing to decoupling of heavy particles. In principle, it
is possible to write down a lot of different EL describing effects
of new physics beyond the SM. In Ref. \cite{2} the EL generated at
a tree level in a general renormalizable gauge theory have been
derived. These objects by construction contain a great number of
arbitrary parameters responsible for specific processes. But it is
well known that a renormalizable theory includes a small number of
independent constants due to relations between them. The
renormalizability of the theory is resulted in the renormalization
group (RG) equations for scattering amplitudes \cite{3}. In Ref.
\cite{4} it has been proven that RG equation can be used to obtain
a set of relations between the parameters of the EL. Two main
observations were used. First, it has been shown that a heavy
virtual state may be considered as an external field scattering SM
light particles. Second, the renormalization of the vertices,
describing scattering on the external field, can be determined by
the $\beta $- and $\gamma $- functions calculated with light
particles, only. Hence, the relations mentioned above follow. As
an example the SM with the heavy Higgs scalar has been
investigated. In the decoupling region the RG equations for
scattering amplitudes have been reduced to the ones for vertices
describing the scattering of light particles on the external field
substituting the corresponding virtual heavy field. In
Ref.\cite{4} the only scalar field of the theory was taken as the
heavy particle, and no mixing between the heavy and the light
fields at the one-loop level has been considered. Here, we are
going to investigate the Yukawa model with a heavy scalar field
$\chi$ and a light scalar field $\varphi$. The purposes of our
investigation are two fold: to derive the one-loop RG relation for
the four-fermion scattering amplitude in the decoupling region and
to find out the possibility of reducing this relation in the
equation for vertex describing the scattering of light particles
on the external field when the mixing between heavy and light
virtual states takes place. In Ref.\cite{4} the specific algebraic
identities originated from the RG equation for scattering
amplitude have been derived. When the explicit couplings in EL are
unknown and represented by the arbitrary parameters, one may treat
the identities as the equations dependent on the parameters and
appropriate $\beta -$ and $\gamma -$ functions. If due to a
symmetry the number of $\beta -$ and $\gamma -$ functions is less
than the number of RG relations, one can obtain non trivial system
of equations for the parameters mentioned. This was shown for the
gauge couplings \cite{4}. In present paper we derive RG relations
for the EL parameters in the model including one-loop mixing of
heavy and light fields.

\section{Renormalization group relation for amplitude}

\begin{figure}[t]
\begin{center}
  \epsfxsize=0.4\textwidth
  \epsfbox[0 0 600 600]{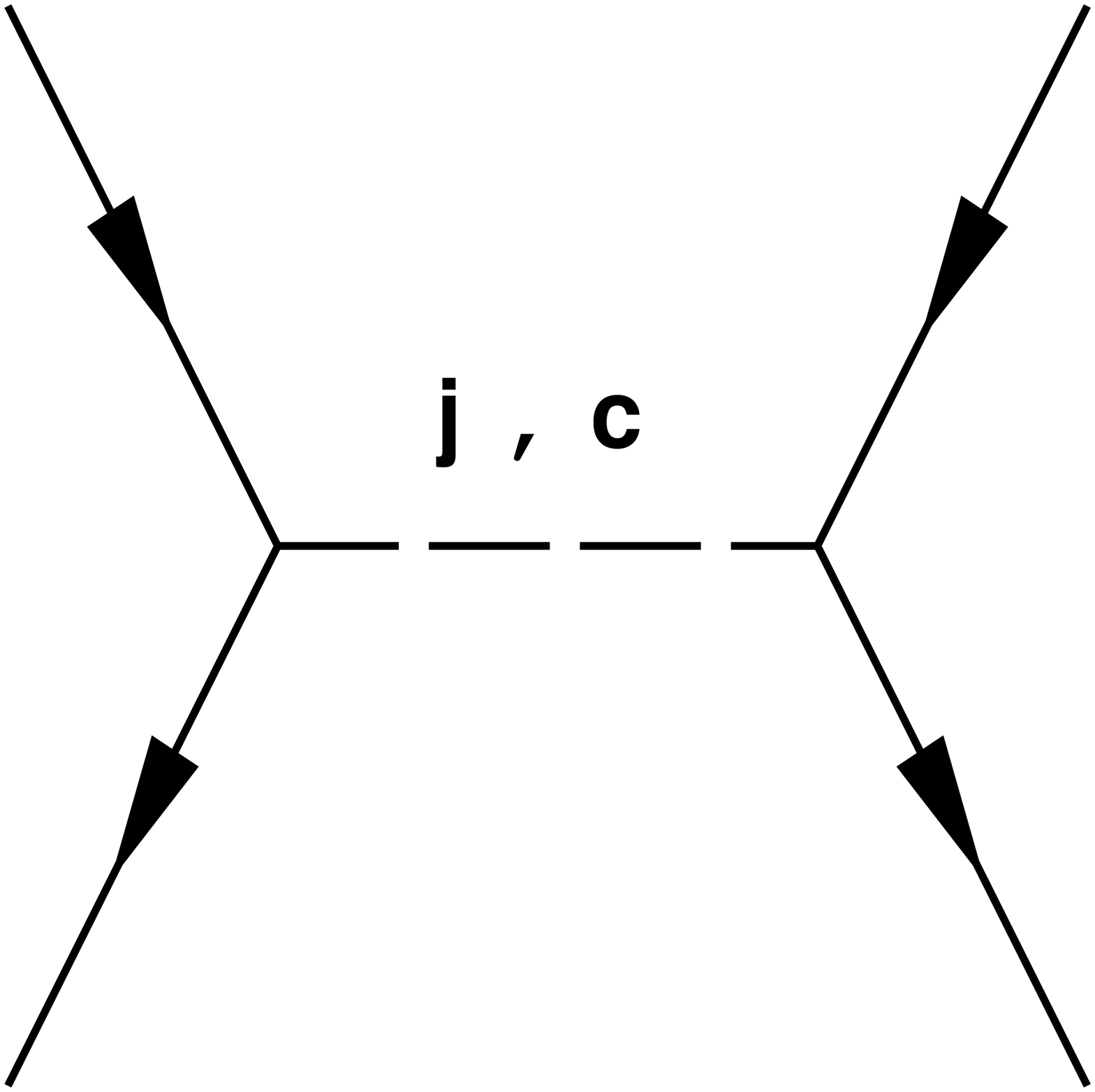}
  \caption
{
   Tree level contribution to the four-fermion amplitude.
}
\label{fig:tree}
\end{center}
\end{figure}

\begin{figure}[t]
\begin{center}
  \epsfxsize=0.9\textwidth
  \epsfbox[0 0 1800 600]{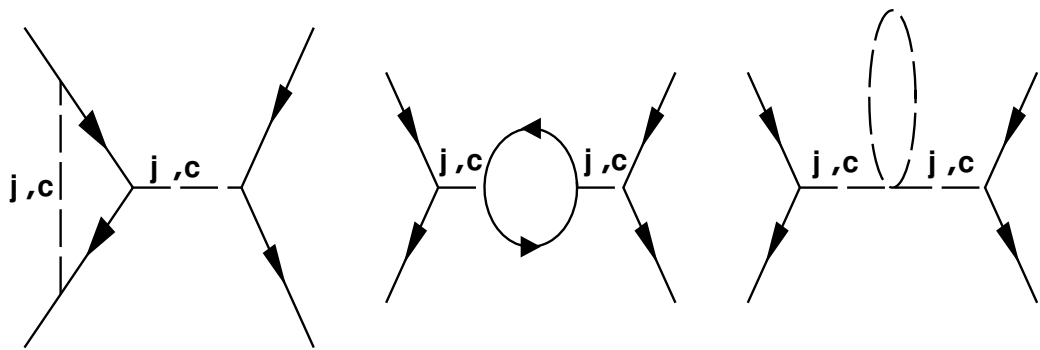}
  \caption
{
   One-loop level contribution to the one-particle reducible four-fermion amplitude.
}
\label{fig:loop}
\end{center}
\end{figure}

The Lagrangian of the model reads
\begin{eqnarray}\label{1}
{\cal L}&=&\frac{1}{2}{\left( \partial_{\mu}\varphi \right) }^{2}-
\frac{m^{2}}{2}{\varphi}^{2}-\lambda{\varphi}^{4}+ \frac{1}{2}{\left(
\partial_{\mu}\chi \right) }^{2}- \frac{{\Lambda}^{2}}{2}{\chi}^{2}-
\xi{\chi}^{4}+\nonumber \\
&&\rho {\varphi}^{2}{\chi}^{2}+{\bar\psi}\left(
i\partial_{\mu}\gamma_{\mu}-M-G_{\varphi}\varphi- G_{\chi}\chi \right) \psi,
\end{eqnarray}
where $\psi$ is a Dirac spinor field.
The $S$-matrix element for the four-fermion scattering at the one-loop level
is given by
\begin{eqnarray}\label{2}
{\hat S}&=&-\frac{i}{2}\int\frac{dp_{1}}{{\left( 2\pi\right)
}^{4}}...\frac{dp_{4}}{{\left( 2\pi\right) }^{4}}{\left( 2\pi\right) }^{4}\delta\left(
p_{1}+...+p_{4}\right) {\cal N}\left[ S_{1PR}+ S_{box}\right],\nonumber \\
S_{1PR}&=&\sum\limits_{{\phi}_{1},{\phi}_{2}=\varphi,
\chi}G_{{\phi}_{1}} G_{{\phi}_{2}}\left(
\frac{{\delta}_{{\phi}_{1}{\phi}_{2}}}{s-m_{{\phi}_{1}}}+\frac{1}{s-
m_{{\phi}_{1}}}{\Pi}_{{\phi}_{1}{\phi}_{2}}\left( s \right) \frac{1}{s-
m_{{\phi}_{2}}}\right) \times\nonumber \\
&& {\bar\psi}\left( p_{1}\right) \left(1+2\Gamma\left( p_{2},
-p_{1}- p_{2}\right)  \right) \psi\left( p_{2}\right)\times {\bar\psi}\left(
p_{4}\right) \psi\left( p_{3}\right),
\end{eqnarray}
where $s={\left( p_{1}+p_{2}\right) }^{2}$, $S_{1PR}$ is the
contribution from the one-particle reducible diagrams shown in the
Figs.\ref{fig:tree}-\ref{fig:loop} and $S_{box}$ is the
contribution from the box diagram. The one-loop polarization
operator of scalar fields ${\Pi}_{{\phi}_{1}{\phi}_{2}}$ and the
one-loop vertex function $\Gamma$ are usually defined trough the
Green functions:
\begin{eqnarray}\label{3}
D_{{\phi}_{1}{\phi}_{2}}\left( s \right)
&=&\frac{{\delta}_{{\phi}_{1}{\phi}_{2}}}{s-m_{{\phi}_{1}}}+\frac{1}{s-
m_{{\phi}_{1}}}{\Pi}_{{\phi}_{1}{\phi}_{2}}\left( s \right) \frac{1}{s-
m_{{\phi}_{2}}},\nonumber \\
G_{\phi\phi\psi}\left( p,q \right)&=&-\sum\limits_{{\phi}_{1}
}G_{{\phi}_{1}} D_{{\phi}_{1}\phi}\left( q^{2} \right) S_{\psi}\left( p \right)
\left(1+\Gamma\left( p, q\right)  \right) S_{\psi} \left( -p-q\right),
\end{eqnarray}
where $ S_{\psi}$ is the spinor propagator in the momentum representation. The
renormalized fields, masses and charges are defined as follows
\begin{eqnarray}\label{4}
\psi &=&Z_{\psi}^{-1/2}{\psi}_{0},\quad\left( \begin{array}{c}\varphi \\ \chi
\end{array}\right)= Z_{\phi}^{-1/2}\left( \begin{array}{c}{\varphi}_{0} \\
{\chi}_{0} \end{array}\right),\quad \left( \begin{array}{c}\Gvar \\ \Gchi
\end{array}\right)= Z_{G}^{-1}\left( \begin{array}{c}G_{{\varphi}_{0}} \\
G_{{\chi}_{0}} \end{array}\right),\nonumber \\
M^{2}&=& M_{0}^{2}-{\delta M}^{2},\qquad\quad m^{2}= m_{0}^{2}-{\delta
m}^{2},\qquad\quad {\Lambda}^{2}= {\Lambda}_{0}^{2}-{\delta \Lambda}^{2},
\end{eqnarray}

Using the dimensional regularization (the dimension of the momentum space is
$D=4-\varepsilon$) and the $\overline{MS}$ renormalization scheme \cite{5} one can
compute the renormalization constants
\begin{eqnarray}\label{5}
Z_{\psi}&=&1-\frac{1}{16{\pi}^{2}\varepsilon}\left( {\Gvar}^{2}+
{\Gchi}^{2}\right) ,\quad {\delta M}^{2}=\frac{3}{8{\pi}^{2}\varepsilon}\left(
{\Gvar}^{2}+ {\Gchi}^{2}\right) M^{2},\nonumber \\
Z_{\phi}^{1/2}&=&1-\frac{1}{8{\pi}^{2}\varepsilon}\left(
\begin{array}{cc}{\Gvar}^{2}&
2\Gvar\Gchi\frac{{\Lambda}^{2}-6 M^{2}}{{\Lambda}^{2}-
m^{2}}\\-2\Gvar\Gchi\frac{m^{2}-6 M^{2}}{{\Lambda}^{2}-
m^{2}}& {\Gchi}^{2}\end{array}\right) , \nonumber \\
{\delta m}^{2}&=&\frac{1}{4{\pi}^{2}\varepsilon}\left( \left(
{\Gvar}^{2}+6\lambda\right) m^{2}-6 {\Gvar}^{2}M^{2}-
\rho{\Lambda}^{2}\right),\nonumber \\
{\delta \Lambda}^{2}&=&\frac{1}{4{\pi}^{2}\varepsilon}\left( \left(
{\Gchi}^{2}+6\xi\right) {\Lambda}^{2}-6 {\Gchi}^{2}M^{2}-\rho
m^{2}\right),\nonumber \\
Z_{G}^{-1}&=&\left[ 1-\frac{3}{16{\pi}^{2}\varepsilon}\left(
G_{\varphi}^{2}+ G_{\chi}^{2}\right) \right] {\left( Z_{\phi}^{1/2}\right) }^{T}.
\end{eqnarray}
From Eq.(\ref{5}) we obtain the appropriate $\beta -$ and $\gamma -$ functions
\cite{5} at the one-loop level:
\begin{eqnarray}\label{6}
&& \betavar =\frac{d\Gvar}{d ln \kappa}=\frac{1}{16{\pi}^{2}}\left(
5{\Gvar}^{3}+3\Gvar {\Gchi}^{2}-4\frac{m^{2}-6 M^{2}}{{\Lambda}^{2}-
m^{2}}\Gvar {\Gchi}^{2} \right) ,\nonumber \\
&& \betachi =\frac{d\Gchi}{d ln \kappa}=\frac{1}{16{\pi}^{2}}\left(
5{\Gchi}^{3}+3\Gchi {\Gvar}^{2}+4\frac{{\Lambda}^{2}-6
M^{2}}{{\Lambda}^{2}-m^{2}}\Gchi {\Gvar}^{2} \right) ,\nonumber \\
&& \gammam =-\frac{d ln m^{2}}{d ln \kappa}=-\frac{1}{4{\pi}^{2}}\left(
{\Gvar}^{2}\frac{m^{2}-6 M^{2}}{m^{2}}+6\lambda-\rho
\frac{{\Lambda}^{2}}{m^{2}} \right) ,\nonumber \\
&& \gammaL =-\frac{d ln {\Lambda}^{2}}{d ln \kappa}=-
\frac{1}{4{\pi}^{2}}\left( {\Gchi}^{2}\left( 1-
6\frac{M^{2}}{{\Lambda}^{2}}\right) +6\xi-\rho
\frac{m^{2}}{{\Lambda}^{2}} \right) ,\nonumber \\
&& \gammapsi =-\frac{d ln \psi}{d ln
\kappa}=\frac{1}{32{\pi}^{2}}\left({\Gvar}^{2}+{\Gchi}^{2} \right) .
\end{eqnarray}
Then, the $S$-matrix element can be expressed in terms of the renormalized
quantities (\ref{4}). The contribution from the one-particle reducible diagrams becomes
\begin{eqnarray}\label{7}
S_{1PR}&=&\sum\limits_{{\phi}_{1},{\phi}_{2}}G_{{\phi}_{1}}
G_{{\phi}_{2}}\left( \frac{{\delta}_{{\phi}_{1}{\phi}_{2}}}{s-
m_{{\phi}_{1}}}+\frac{1}{s-
m_{{\phi}_{1}}}{\Pi}_{{\phi}_{1}{\phi}_{2}}^{fin}\left( s \right) \frac{1}{s-
m_{{\phi}_{2}}}\right) \nonumber \\
&& {\bar\psi}\left( p_{1}\right)
\left(1+2{\Gamma}^{fin}\left( p_{2}, -p_{1}- p_{2}\right)  \right) \psi\left(
p_{2}\right)\times {\bar\psi}\left( p_{4}\right) \psi\left( p_{3}\right),
\end{eqnarray}
where the functions ${\Pi}_{{\phi}_{1}{\phi}_{2}}^{fin}$ and
${\Gamma}^{fin}$ are the expressions ${\Pi}_{{\phi}_{1}{\phi}_{2}}$ and
$\Gamma$ without the terms proportional to $1/\varepsilon$. Since the quantity
$S_{box}$ is finite, the renormalization leaves it without changes.
Introducing the RG operator at the one-loop level \cite{6}
\begin{eqnarray}\label{8}
{\cal D}&=&\frac{d}{d ln \kappa}=\frac{\partial}{\partial ln \kappa}+{\cal
D}^{(1)}=\nonumber\\
&&\frac{\partial}{\partial ln \kappa}+\sum\limits_{\phi}
\betaphi\frac{\partial}{\partial \Gphi}-\gammam \frac{\partial}{\partial ln m^{2}}-\gammaL
\frac{\partial}{\partial ln {\Lambda}^{2}}-\gammapsi \frac{\partial}{\partial ln \psi}
\end{eqnarray}
we determine that the following relation holds for the $S$-matrix element
\begin{equation}\label{9}
{\cal D}\left( S_{1PR}+S_{box} \right) =\frac{\partial
S_{1PR}^{(1)}}{\partial ln \kappa}+{\cal D}^{(1)} S_{1PR}^{(0)}=0,
\end{equation}
where the $S_{1PR}^{(0)}$ and the $S_{1PR}^{(1)}$ are the contributions to the
$S_{1PR}$ at the tree level and at the one-loop level, respectively:
\begin{equation}\label{10}
S_{1PR}^{(0)}=\left( \frac{{\Gvar}^{2}}{s-m^{2}}+\frac{{\Gchi}^{2}}{s-
{\Lambda}^{2}} \right) {\bar\psi}\psi\times{\bar\psi}\psi ,
\end{equation}
\begin{eqnarray}\label{11}
\frac{\partial S_{1PR}^{(1)}}{ \partial ln
\kappa}&=&\frac{{\bar\psi}\psi\times{\bar\psi}\psi }{4{\pi}^{2}}\left( -\left(
{\Gvar}^{2}+{\Gchi}^{2}\right) \left( \frac{{\Gvar}^{2}}{s-
m^{2}}+\frac{{\Gchi}^{2}}{s-{\Lambda}^{2}} \right) + \right. \nonumber \\
&& \frac{{\Gvar}^{2}\left( \rho{\Lambda}^{2}-6\lambda
m^{2}+{\Gvar}^{2}\left( 6M^{2}-s\right) \right) }{{\left( s-m^{2}\right) }^{2}}+
\frac{2{\Gvar}^{2}{\Gchi}^{2}\left( 6M^{2}-s\right) }{\left( s-m^{2}\right)
\left( s-{\Lambda}^{2}\right) }+\nonumber \\
&&\left. \frac{{\Gchi}^{2}\left( \rho m^{2}-6\xi{\Lambda}^{2}+{\Gchi}^{2}\left(
6M^{2}-s\right) \right) }{{\left( s-{\Lambda}^{2}\right) }^{2}} \right) .
\end{eqnarray}

The first term in Eq.(\ref{11}) is originated from the one-loop
correction to the fermion-scalar vertex. The rest terms are
connected with the polarization operator of scalars. The third
term describes the one-loop mixing between the scalar fields. It
is canceled in the RG relation (\ref{9}) by the mass-dependent
terms in the $\beta -$ functions produced by the non-diagonal
elements in $Z_{\phi}$. Eq.(\ref{9}) is the consequence of the
renormalizability of the model. It insures the leading logarithm
terms of the one-loop $S$-matrix element to reproduce the
appropriate tree-level structure. In contrast to the familiar
treatment we are not going to improve scattering amplitudes by
solving Eq.(\ref{9}). We will use it as an algebraic identity
implemented in the renormalizable theory. Naturally if one knows
the explicit couplings expressed in terms of the basic set of
parameters of the model, this RG relation is trivially fulfilled.
But the situation changes when the couplings are represented by
unknown arbitrary parameters as it occurs in the EL approach
\cite{1},\cite{2}. In this case the RG relations are the algebraic
equations dependent on these parameters and appropriate $\beta -$
and $\gamma -$ functions. In the presence of a symmetry the number
of $\beta -$ and $\gamma -$ functions is less than the number of
RG relations. So, one has non trivial system of equations relating
the parameters of EL. Such a scenario is realized for the gauge
coupling as it has been demonstrated in \cite{4}. Although the
considered simple model has no gauge couplings and no relation
between the EL parameters occurs, we are able to demonstrate the
general procedure of deriving the RG relations for EL parameters
in the theory with one-loop mixing. This is essential for dealing
with the EL describing deviations from the SM. At energies $s\ll
{\Lambda}^{2}$ the heavy scalar field $\chi$ is decoupled. So, the
four-fermion scattering amplitude consists of the contribution of
the model with no heavy field $\chi$ plus terms of the order $s/
{\Lambda}^{2}$. The expansion of the heavy scalar propagator
\begin{equation}\label{12}
\frac{1}{s-{\Lambda}^{2}}\to -
\frac{1}{{\Lambda}^{2}}\left(1+O\left(\frac{s}{{\Lambda}^{2}}\right) \right)
\end{equation}
in Eq.(\ref{10}) is resulted in the effective contact four-fermion interaction
\begin{equation}\label{13}
{\cal L}_{eff}=-\alpha{\bar\psi}\psi\times{\bar\psi}\psi ,\quad\alpha=
\frac{{\Gchi}^{2}}{{\Lambda}^{2}},
\end{equation}
and the tree level contribution to the amplitude becomes
\begin{equation}\label{14}
S_{1PR}^{(0)}=\left( \frac{{\Gvar}^{2}}{s-m^{2}}-
\alpha+O\left(\frac{s}{{\Lambda}^{4}}\right) \right)
{\bar\psi}\psi\times{\bar\psi}\psi .
\end{equation}

In the decoupling region the lowest order effects of the heavy
scalar are described by the parameter $\alpha$, only. The method
of constructing the RG equation in terms of the low energy
quantities $G_{\varphi}, \lambda, m, M, \alpha$ was proposed in
\cite{6}. As it has been demonstrated in \cite{6}, the
redefinition of the parameters of the model allows to remove all
the heavy particle loop contributions to Eq.(\ref{11}). Let us
define a new set of fields, charges and masses $\tpsi$, $\tGvar$,
$\tGchi$, $\tL$, $\tm$, $\tM$
\begin{eqnarray}\label{15}
&& {\Gvar}^{2}={\tGvar}^{2}\left( 1+\frac{3{\tGchi}^{2}}{16{\pi}^{2}}
ln\frac{{\kappa}^{2}}{{\tL}^{2}} \right) ,
\quad{\Gchi}^{2}={\tGchi}^{2}\left( 1+\frac{3{\tGchi}^{2}}{16{\pi}^{2}}
ln\frac{{\kappa}^{2}}{{\tL}^{2}} \right) , \nonumber \\
&& m^{2} ={\tm}^{2}\left( 1-\frac{\tilde\rho}{8{\pi}^{2}}
\frac{{\tL}^{2}}{{\tm}^{2}} ln\frac{{\kappa}^{2}}{{\tL}^{2}} \right) ,
\quad{\Lambda}^{2}={\tL}^{2}\left( 1+\frac{3\tilde\xi}{4{\pi}^{2}}
ln\frac{{\kappa}^{2}}{{\tL}^{2}} \right) ,\nonumber \\
&& \psi=\tpsi \left( 1-\frac{{\tGchi}^{2}}{64{\pi}^{2}}
ln\frac{{\kappa}^{2}}{{\tL}^{2}} \right) .
\end{eqnarray}
One is able to rewrite the differential operator (\ref{8}) in
terms of these new low-energy parameters:
\begin{equation}\label{16}
{\cal D}=\frac{\partial}{\partial ln \kappa}+{\tilde{\cal D}}^{(1)}=
\frac{\partial}{\partial ln \kappa}+\sum\limits_{\phi} \tbetaphi\frac{\partial}{\partial
\tGphi}-\tgammam \frac{\partial}{\partial ln {\tm}^{2}}-\tgammaL \frac{\partial}{\partial ln
{\tL}^{2}}-\tgammapsi \frac{\partial}{\partial ln \tpsi}
\end{equation}
where ${\tilde\beta}-$ and ${\tilde\gamma}-$ functions are obtained from the one-
loop relations (\ref{6}) and (\ref{15})
\begin{eqnarray}\label{17}
&& \tbetavar =\frac{1}{16{\pi}^{2}}\left( 5{\tGvar}^{3}-4\frac{{\tm}^{2}-6
{\tM}^{2}}{{\tL}^{2}-{\tm}^{2}}\tGvar {\tGchi}^{2} \right) ,\nonumber \\
&& \tbetachi =\frac{1}{16{\pi}^{2}}\left( 2{\tGchi}^{3}+\left( 3+
4\frac{{\tL}^{2}-6 {\tM}^{2}}{{\tL}^{2}-{\tm}^{2}}\right) \tGchi {\tGvar}^{2}
\right) ,\nonumber \\
&& \tgammam =-\frac{1}{4{\pi}^{2}}\left( {\tGvar}^{2}\frac{{\tm}^{2}-6
{\tM}^{2}}{{\tm}^{2}}+6{\tilde\lambda} \right) ,\nonumber \\
&& \tgammaL =-\frac{1}{4{\pi}^{2}}\left( {\tGchi}^{2}\left( 1-
6\frac{{\tM}^{2}}{{\tL}^{2}}\right) -{\tilde\rho}\frac{{\tm}^{2}}{{\tL}^{2}}
\right) ,\nonumber \\
&& \tgammapsi =\frac{1}{32{\pi}^{2}}{\tGvar}^{2}.
\end{eqnarray}
Hence, one immediately notices that ${\tilde\beta}-$ and
${\tilde\gamma}-$ functions contain only the light particle loop
contributions, and all the heavy particle loop terms are
completely removed from them. The $S$-matrix element expressed in
terms of new parameters satisfies the following RG relation
\begin{equation}\label{18}
{\cal D}\left( S_{1PR}+S_{box} \right) =\frac{\partial {\tilde
S}_{1PR}^{(1)}}{\partial ln \kappa}+{\tilde{\cal D}}^{(1)} {\tilde
S}_{1PR}^{(0)}=0,
\end{equation}
\begin{equation}\label{19}
{\tilde S}_{1PR}^{(0)}=\left( \frac{{\tGvar}^{2}}{s-{\tm}^{2}}-
{\tilde\alpha}+O\left( \frac{s^{2}}{{\tL}^{4}} \right) \right)
{\bar\tpsi}\tpsi\times{\bar\tpsi}\tpsi ,
\end{equation}
\begin{eqnarray}\label{20}
\frac{\partial {\tilde S}_{1PR}^{(1)}}{ \partial ln
\kappa}&=&\frac{{\bar\tpsi}\tpsi\times{\bar\tpsi}\tpsi }{4{\pi}^{2}}\left( -
\frac{{\tGvar}^{4}}{s-{\tm}^{2}}+\frac{{\tGvar}^{2}\left( -
6{\tilde\lambda}{\tm}^{2}+{\tGvar}^{2}\left( 6{\tM}^{2}-s\right) \right) }{{\left(
s-{\tm}^{2}\right) }^{2}}+\right. \nonumber \\
&&\left. {\tilde\alpha}{\tGvar}^{2}-
\frac{2{\tGvar}^{2}{\tilde\alpha}\left( 6{\tM}^{2}-s\right) }{s-m^{2}}+
O\left( \frac{s^{2}}{{\tL}^{4}} \right) \right) ,
\end{eqnarray}
where ${\tilde\alpha}={\tGchi}^{2}/{\tL}^{2}$ is the redefined effective four-
fermion coupling. As one can see, Eq.(\ref{20}) includes all the terms of Eq.(\ref{11}) except
for the heavy particle loop contributions. It depends on the low energy quantities
$\tpsi$, $\tGvar$, $\tilde\alpha$, $\tilde\lambda$, $\tm$, $\tM$. The first and the
second terms in Eq.(\ref{20}) are just the one-loop amplitude calculated within the
model with no heavy particles. The third and the fourth terms describe the light
particle loop correction to the effective four-fermion coupling and the mixing of
heavy and light virtual fields.

\section{Elimination of one-loop scalar field mixing}

Due to the mixing term it is impossible to split the RG relation
(\ref{18}) for the S- matrix element into the one for vertices.
Hence, we are not able to consider Eq.(\ref{18}) in the framework
of the scattering of light particles on an external field induced
by the heavy virtual scalar as it has been done in \cite{4}. But
this is an important step in deriving the RG relation for EL
parameters. Fortunately, there is a simple procedure allowing to
avoid the mixing in Eq.(\ref{20}). The way is to diagonalize the
leading logarithm terms of the scalar polarization operator in the
redefinition of the $\tilde\varphi$, $\tilde\chi$, $\tGvar$,
$\tGchi$
\begin{eqnarray}\label{21}
&&\left( \begin{array}{c}\varphi \\ \chi \end{array}\right)=
{\zeta}^{1/2}\left( \begin{array}{c}{\tilde\varphi} \\ {\tilde\chi}
\end{array}\right),\quad \left( \begin{array}{c}G_{\varphi} \\ G_{\chi}
\end{array}\right)=\left[ 1+\frac{3{\tGchi}^{2}}{32{\pi}^{2}}
ln\frac{{\kappa}^{2}}{{\tL}^{2}} \right]
{\left( {\zeta}^{-1/2}\right) }^{T}\left(
\begin{array}{c}\tGvar \\ \tGchi \end{array}\right)
,\nonumber \\
&& {\zeta}^{1/2}=1- \frac{\tGvar\tGchi}{8{\pi}^{2} \left({\tL}^{2}-
{\tm}^{2}\right) }ln\frac{{\kappa}^{2}}{{\tL}^{2}}\left( \begin{array}{cc}0 &
{\tL}^{2}-6 {\tM}^{2}\\-{\tm}^{2}-6 {\tM}^{2}& 0\end{array}\right) .
\end{eqnarray}
The appropriate ${\tilde\beta}-$ functions
\begin{equation}\label{22}
 \tbetavar =\frac{5{\tGvar}^{3}}{16{\pi}^{2}},\quad
 \tbetachi =\frac{1}{16{\pi}^{2}}\left( 2{\tGchi}^{3}+3\tGchi {\tGvar}^{2}
\right)
\end{equation}
contain no terms connected with mixing between light and heavy scalars. So, the
fourth term in Eq.(\ref{20}) is removed, and the RG relation for the $S$-matrix element
becomes
\begin{equation}\label{23}
{\cal D}\left( S_{1PR}+S_{box} \right) =\frac{\partial {\tilde
S}_{1PR}^{(1)}}{\partial ln \kappa}+{\tilde{\cal D}}^{(1)} {\tilde
S}_{1PR}^{(0)}=0,
\end{equation}
\begin{equation}\label{24}
{\tilde S}_{1PR}^{(0)}=\left( \frac{{\tGvar}^{2}}{s-{\tm}^{2}}-
{\tilde\alpha}+O\left( \frac{s^{2}}{{\tL}^{4}} \right) \right)
{\bar\tpsi}\tpsi\times{\bar\tpsi}\tpsi ,
\end{equation}
\begin{eqnarray}\label{25}
\frac{\partial {\tilde S}_{1PR}^{(1)}}{ \partial ln
\kappa}&=&\frac{{\bar\tpsi}\tpsi\times{\bar\tpsi}\tpsi }{4{\pi}^{2}}\left( -
\frac{{\tGvar}^{4}}{s-{\tm}^{2}}+\frac{{\tGvar}^{2}\left( -
6{\tilde\lambda}{\tm}^{2}+{\tGvar}^{2}\left( 6{\tM}^{2}-s\right) \right) }{{\left(
s-{\tm}^{2}\right) }^{2}}+\right. \nonumber \\
&&\left. {\tilde\alpha}{\tGvar}^{2}+
O\left( \frac{s^{2}}{{\tL}^{4}} \right) \right) .
\end{eqnarray}

At ${\tilde\alpha}=0$ Eq.(\ref{23}) is just the RG identity for the scattering amplitude
calculated in the absence of the heavy particles. The terms of order $\tilde\alpha$
describe the RG relation for the effective low-energy four-fermion interaction in the
decoupling region. The last one can be reduced in the RG relation for the vertex
describing the scattering of the light particle (fermion) on the external field
$\sqrt{\tilde\alpha}$ substituting the virtual heavy scalar:
\begin{equation}\label{26}
{\cal D}\left( \sqrt{\tilde\alpha}{\bar\tpsi}\tpsi \right)
=\frac{{\tGvar}^{2}}{8{\pi}^{2}}\sqrt{\tilde\alpha}{\bar\tpsi}\tpsi+{\tilde{\cal
D}}^{(1)} \left( \sqrt{\tilde\alpha}{\bar\tpsi}\tpsi \right) =0,
\end{equation}
where
\begin{eqnarray}\label{27}
&&{\tilde{\cal D}}^{(1)}= \tbetavar\frac{\partial}{\partial \tGvar}-
{\tilde\gamma}_{\alpha}\frac{\partial}{\partial ln {\tilde\alpha}}-\tgammam
\frac{\partial}{\partial ln {\tm}^{2}}-\tgammapsi \frac{\partial}{\partial ln
\tpsi},\nonumber \\
&& {\tilde\gamma}_{\alpha}=-{\cal D}{\tilde\alpha}=-
\frac{1}{8{\pi}^{2}}\left( 3{\tGvar}^{2}+O\left({\tilde\alpha}\right) \right) .
\end{eqnarray}

Eqs.(\ref{23})-(\ref{27}) is the main result of our investigation.
One can derive them with only the knowledge about the EL
(\ref{13}) and the Lagrangian of the model with no heavy
particles. One also has to ignore all the heavy particle loop
contributions to the RG relation and the one-loop mixing between
the heavy and the light fields. Eqs.(\ref{23})-(\ref{27}) depend
on the effective low-energy parameters, only. But as the
difference between the original set of parameters and the
low-energy one is of one-loop order, one may freely substitute
them in Eqs.(\ref{23})-(\ref{26}).

\section{Discussion}

Let us discuss the results obtained. The RG relation for the
four-fermion scattering amplitude is derived in the decoupling
region $s\ll {\Lambda}^{2}$. It was shown that one can redefine
the parameters and the fields of the model in order to remove all
the heavy particle loop contributions to the RG relation. Then the
RG relation becomes dependent on the low-energy physics
parameters, only. As the RG operator coefficients and the
difference between the original parameters and the redefined ones
are of the one-loop order one can substitute one set of parameters
by another at the lowest level. Thus, we extend the result of
Ref.\cite{4} to the case when mixing terms are present. The
additional transformation of fields and charges allows one to
diagonalize the leading logarithm terms of the scalar polarization
operator and to avoid the contributions to the RG relation
originated from the one-loop mixing between heavy and light field.
Since the difference between the diagonalized fields and charges
and the original ones is of one-loop order, one may simply omit
one-loop mixing terms in the RG relation at the lower level. Then
it is possible to reduce the RG relation for $S$-matrix element to
the one for vertex describing the scattering of light particles on
the external field induced by the heavy virtual particle. In fact,
this result is independent on the specific features of the
considered model, as it was shown in \cite{4}.

The RG relations of the considered type may be used in searching
for the dependences between the parameters of EL describing
physics beyond the SM. For example, let a symmetry requires the
same charge structure for some effective Lagrangians. Then the
number of unknown ${\tilde\beta}-$ and ${\tilde\gamma}-$ functions
is less than the number of RG relations, and it is possible to
derive non-trivial solutions for the parameters. The present
results allow to omit the one-loop mixing diagrams in construction
of the RG relations for the tree-level EL.

\newpage

\clearpage

\newpage
\clearpage

\end{document}